\documentclass[a4paper,twoside]{article}

\usepackage{epsfig}
\usepackage{subcaption}
\usepackage{calc}
\usepackage{amssymb}
\usepackage{amstext}
\usepackage{amsmath}
\usepackage{amsthm}
\usepackage{multicol}
\usepackage{pslatex}
\usepackage{apalike}
\usepackage[bottom]{footmisc}
\usepackage{SCITEPRESS}     % Please add other packages that you may need BEFORE the SCITEPRESS.sty package.
\usepackage[hidelinks]{hyperref}

\begin{document}

\title{A Next-Generation Digital Procurement Workspace Focusing on Information Integration, Automation, Analytics, and Sustainability}

\author{\authorname{Jan-David Stütz\sup{1}, Oliver Karras\sup{2}, Allard Oelen\sup{2}, and Sören Auer\sup{2}}
 \affiliation{\sup{1}Robert Bosch GmbH, Stuttgart, Germany}
 \affiliation{\sup{2}TIB - Leibniz Information Centre for Science and Technology, Hannover, Germany}
 \email{jan-david.stuetz@de.bosch.com, \{oliver.karras, allard.oelen, soeren.auer\}@tib.eu}
 }


\abstract{Recent events such as wars, sanctions, pandemics, and climate change have shown the importance of proper supply network management.
A key step in managing supply networks is procurement.
We present an approach for realizing a next-generation procurement workspace that aims to facilitate resilience and sustainability.
To achieve this, the approach encompasses a novel way of information integration, automation tools as well as analytical techniques.
As a result, the procurement can be viewed from the perspective of the environmental impact, comprising and aggregating sustainability scores along the supply chain.
We suggest and present an implementation of our approach, which is meanwhile used in a global Fortune 500 company.
We further present the results of an empirical evaluation study, where we performed in-depth interviews with the stakeholders of the novel procurement platform to validate its adequacy, usability, and innovativeness.
}

\onecolumn \maketitle \normalsize \setcounter{footnote}{0} \vfill

\section{Introduction}\label{sec1}
Today (2023), twenty-seven years are left until the goals of the Paris Agreement have to be reached. 
Facilitating resilience and sustainability in supply networks in procurement can contribute to accomplishing the set goals. 
Recent events, such as wars, sanctions, pandemics, and catastrophes, show even more the crucial role of a resilient and sustainable supply chain. 
Some known supply chains that were considered stable are now disrupted. 
Failing suppliers cannot be used any longer and must be exchanged with new, more stable, or environmentally friendlier ones. 
Current inflation pressures procurement and supply chain management (SCM) even more. 
Prices and product quality should ideally be stable, which can be a difficult challenge since Europe's raw materials and energy costs are rising~\cite{ari2022surging}. 

At the Robert Bosch GmbH (Bosch), there is currently no central digital procurement tool that combines data silos and provides valuable features in the day-to-day work of purchasers regarding information integration, automation, analytics, and sustainability. 
Various tools are used alongside the entire procurement process, which increases the complexity even for simple tasks. 
Purchasers have to pull external and internal information.
Automation of simple and low-risk tasks and analytics of various supply chain-related data are rarely present, while sustainability is sometimes completely lacking in existing tools. 

Well-known Enterprise Resource Planning (ERP) systems and specialized procurement applications provide solutions. 
However, as Pekša and Grabis~\cite{Grabis2018} argue, customizing ERP systems are cost and time intensive.
Particularly in the face of rapidly occurring supplier or entire supply chain failures, rapid development and adaptation of these critical systems is essential for driving competitive advantage and improving business performance. 
In addition, existing day-to-day operations involve complex manual processes, lack collaboration, are error-prone and time-consuming, do not provide new insights, and do not support decision-making~\cite{Grabis2018,Tarigan2021}.

In this work, we propose a customized Digital Procurement Workspace (DPW) that includes a Procurement Information Space (PIS), a Procurement Automation and Analytics Space (PAAS), and a Sustainable Sourcing Space (SSS). 
The application and spaces aim to facilitate a faster, more supportive, resilient, and sustainable procurement process. 
The PIS provides personalized, automatically aggregated, and summarized purchasing- and user-relevant information to support purchasers in their daily decisions, to keep them up-to-date, and to allow a quick overview of the most relevant information such as announcements. 
The PAAS provides features to access analyzed or aggregated data. PAAS also highlights key data and allows triggering bots to reduce human-to-human interaction as well as to automize low-risk but time-intensive work. 
While the SSS provides analytical features, it focuses entirely on environmental impact, risk detection, and overall sustainability by, for example, providing a sustainability score. 

We defined overall goals so that the DPW could solve the current problems. Based on those goals, concrete requirements are inferred to ensure that the required functionality is implemented. 
After the implementation, in-depth expert interviews were conducted and analyzed using Mayring's qualitative content analysis to evaluate the developed application, ensure that the defined goals and set requirements were met, and discover novel use cases and benefits. As the interview outcomes indicate, the three key advantages of our approach are increased transparency, efficiency, and decision-making. 
However, this work outlines further dependencies and prerequisites of benefits. 
For example, centralization is a critical prerequisite for increased transparency and collaboration. 
Based on those findings, recommendations going beyond the presentation of the concepts and the developed applications are provided. 

Paper structure: \autoref{relatedWork} presents related work. In \autoref{approach}, we present our approach and its implementation. We present our interview study and its results in \autoref{evaluation}. In \autoref{discussion}, we interpret and discuss our findings before we conclude in \autoref{conclusion}. 

\section{Related Work}\label{relatedWork}
Applications that support procurement are not new: Odoo, Precoro, and Prokuria are a few examples of systems that provide typical procurement functionality.
Besides those tools and some features contained in ERP systems, there is also an increasing rise of literature regarding SCM, public (green) procurement, and different technologies supporting those topics.

\subsection{SCM and Procurement}

The majority of legacy ERP systems are heterogeneous systems that different software companies develop.
Ma and Molná~\cite[p. 231]{Ma2019} mention that it is a ``\textit{big challenge for organizations to develop and implement centralized and integrated management systems based on their existing legacy ERP systems to respond to the dynamic business environment with agility}.''
They suggest the usage of ontologies and propose an integration framework based on ontology learning including a basic workflow~\cite{Ma2019}.
Pekša and Grabis~\cite{Grabis2018} reviewed existing research on the decision-making capabilities of ERP systems. 
They identified different approaches for integrating decision-making logic for companies. Since decision-making techniques should be modifiable, scalable, and portable, they suggest decoupling decision-making logic from ERP systems. This decoupling enables the usage of advanced decision-making techniques. 
Implementing decision-making directly into existing ERP systems is seen as inflexible and cost-intensive, leading to competitive disadvantages.

In the last years, many articles regarding SCM and procurement are published. For example, Rejeb et al.~\cite{Rejeb2018} outline the potential of new technologies like Big Data analysis, Automatization and Robotics, IoT, and Blockchain Technology in procurement-related work since the entire supply chain has already and will continue to have technological shifts. 
Nevertheless, the article does not focus on ERP systems. 
However, Tarigan et al.~\cite{Tarigan2021} investigate and show enhanced ERP's impact on firm performance through green SCM, supplier integration, and internal integration. 
The study suggests that a company must improve productivity, efficiency, speed, and services to innovate and survive in the market, which results in upgrading and adjusting the ERP system. 
This suggestion, in turn, leads to customizations at the ERP systems that are regarding Grabis~\cite{Grabis2019} time-consuming and cost-intensive. 
Nevertheless, this study also suggests that fit-gaps (differences between provided functionality of the ERP system and the company's needs) should be closed.
As another gap, literature often indicates a need to consider a supplier's green performance more instead of only focusing on the classical purchasing criteria. 
Igarashi et al.~\cite{Igarashi2015} conceptualize the inclusion of green (in terms of environmental- and climate-friendly) criteria into the supplier selection process in public procurement since purchasers tend to ignore those impacts. 
Including the environmental and climate impacts in the procurement process can be done when all relevant data are gathered and provided to the corresponding purchasers. 
To achieve that, Barrad et al.~\cite{Barrad2020} focus in their study on emerging technologies and concepts and their ability to improve procurement operations. 
It suggests that classical Big Data approaches like analytics and complex event processing can be explored and adapted to already present procurement processes to help reduce costs. 
The work of AlNuaimi et al.~\cite{AlNuaimi2021} also investigates Big Data analytics and procurement. 
Their work determines that digital procurement does not influence green digital procurement but nevertheless has a significant impact on the Big Data analytics capabilities of the company. 
Furthermore, Big Data analytics as a mediator between digital procurement and environmental digital procurement leads to a positive impact on environmental digital procurement. 
In other words, this work describes the impact of Big Data on a purchaser's decision-making process.

A further improvement in a purchaser's performance can be reached with more user-friendly dashboards. 
Magnus and Rudra~\cite{Magnus2019} discovered a gap between theory and practice regarding the availability of a user-friendly dashboard, where the dashboard design is based on principles of cognition. 
The authors claim that dashboards can facilitate transparency and that improved dashboards enhance decision-making in a supply chain. 
They also indicate that if the information is extracted and displayed conducive, users have a faster overview of needed data. 
It should be of interest to companies to create better applications with the goal of user-friendliness since that will enable employees to work more efficiently along the supply chain. 
Another potential obstacle to enabling Big Data analytics is information silos. 
To break down the information silos along the supply chain, Zhan and Tan~\cite{ZHAN2020559} propose, in their work, an integrated infrastructure in order to enhance supply chain performance. 
Even though there are still obstacles left, the work of Handfield et al.~\cite{emerging-procurement-technology} predicts major shifts in the sourcing and supply chain technology environment in the next years. 
As proposed in some other literature, fostering collaboration between logistics and purchasing departments responsible for managing the supply chain makes sense.~\cite{ILIEZUDOR2014337}.
Hecht and Hofbauer~\cite{Hecht2022} expect potentials of up to 15 percent that are not realized in procurement, even though strategic procurement decisions achieve competitive advantages and guarantee a company's success. 

Literature reviews provide a strong outline of the research direction. 
The comparison of literature reviews regarding green procurement done by us~\cite{https://doi.org/10.48366/r287929} using the Open Research Knowledge Graph (ORKG) shows the relevance of the topic for both the private and the public sectors. 
Significantly, the findings of the compared articles indicate the consideration of user behavior and suggest a critical analysis of the current priority for monetary factors versus sustainability while purchasing. 
The articles of Sönnichsen and Clement~\cite{SONNICHSEN2020118901} and Adjei-Bamfo et al.~\cite{ADJEIBAMFO2019189} indicate missing literature and a lack of research on the investigated topic. Appolloni et al.\cite{APPOLLONI2014122}, on the other hand, indicate that an increase in literature is recognizable. 
Since Appolloni et al. \cite{APPOLLONI2014122} considered only the private sector until 2013 and the other two articles only the public sector until 2017 and 2018, it shows either a decrease regarding the topic, the different handling of the corresponding sector done by researchers, or a subjective view on the amount of published articles.
Nevertheless, the work of Masudin et al.~\cite{doi:10.1080/23311916.2022.2119686} presents a peak and further increase of literature regarding green procurement for both the private and the public sectors in 2018.
Considering the indicated increase in literature and the more emerging awareness of society for sustainability, research will probably continue focusing even more on green procurement. \\

To the best of our knowledge, procurement research mainly focuses on public procurement, while the focus on supply chain-related topics often combines topics of ERP systems, their integration, and their impact. 
Researchers often publish studies regarding the impact of new technologies, especially Big Data, in procurement and SCM. 
At the same time, researchers often combine Big Data with work that fosters green and sustainable supply chains. 

In summary, it seems like current research lacks in defining requirements for digital procurement applications, conceptually describing such an application at all, as well as in evaluating novel and innovative applications in industrial settings. 
This is why our work introduces requirements that outline and summarize the set goals. 
We further present concepts based on the specified requirements. 
Current research, for example, the work of AlNuaimi et al., Grabis, and Tarigan et al., also influenced the motivation, concepts, and implementation. 
Evaluating the developed application with in-depth expert interviews and the resulting suggestions and interpretations enrich this work further. 
In the end, the developed DPW should be comparable to other procurement applications and ERP systems. 
Thus a similar comparison as done by Gomez Llanez et al.~\cite{Odoo} can be achieved. Gomez Llanez et al.~\cite{Odoo} compared the ERP systems Odoo and Openbravo while describing the tools in general, their features, and technical details.

\subsection{Procurement and ERP Systems}\label{procurement_apps}
Different software applications support purchasers already during the procurement process. 
Precoro, Odoo, and Prokuria are such well-known and established applications. 
In the following, we briefly introduce these applications to give an overview on the current state of procurement applications. 

\textit{Precoro}'s\footnote{https://precoro.com/} goal for its purchasing tool is to enable transparency and collaboration while eliminating disorder from the procurement process. 
Therefore, the tool focuses on procurement and is meant for the purchasing department of a small to medium-sized company. 
Since the first step of the procurement process is the Request of Quotation (RfQ) the applications entry point is the RfQ-related view. 
There, a section of the created RfQs and all relevant information and actions, like exporting, filtering, searching, and expanding, are provided. 
As soon as an RfQ is approved or rejected, a notification as an email is triggered. 
Through this email, the notified user can directly access the next step of the procurement process, where a user can either place a purchase order or create a Request for Proposals. 
One prominent feature of Precocor is the creation of customized charts, e.g., to show the purchase order based on the supplier or the department.

\textit{Odoo}\footnote{https://odoo.com/} is considered an ERP system, but provides features to support a company's procurement processes. 
When selecting the purchasing view, all relevant details and actions like filtering, grouping, searching, and assigning tasks of created RfQs are displayed. Creating further RfQs and bills is possible, which can then be downloaded as PDFs or shared directly with colleagues. 
In general, Odoo provides fewer features to support the procurement process. However, as an ERP system, it also covers other business areas like human resources or marketing that are irrelevant to this work~\cite{Odoo}.

Another procurement application that allows the management of RfQs and suppliers is \textit{Prokuria}\footnote{https://prokuria.com/}. In comparison to the other applications, it also provides features regarding auctions.
Starting the application leads to a customizable dashboard view that displays different kinds of customizable and interactable so-called widgets, like a table that shows only the currently active RfQs or a table containing all suppliers. 
Prokuria provides different kinds of detailed views, for example, for suppliers and events that contain RfQs and auctions. 
Since suppliers can directly place bids on open and active RfQs, they automatically get a notification via email if they are selected. 
Purchasers can compare the suppliers' bids in a comparison view that offers visual representation in the form of a chart, where a scoring system is provided. 
A report can be viewed and downloaded to analyze the entire generated data.  
Another prominent feature is the possibility of allowing communication between suppliers and purchasers based on the internal message system. \vspace{0.2cm}

The presented applications claim to work well in small to medium-sized companies and only if the entire procurement process is done inside the application. 
Customized features can not be easily added and developed, and a further dependency arises as a company. 
Bosch is a company with distributed departments and already existing solutions for various business cases. For this reason, a transfer to such an application does not make sense.
Furthermore, the presented solutions miss general features or possibilities for customizations regarding information integration, analytics, and automation mainly because some of the company's data is deeply embedded in a technological ecosystem. 
Additionally, aspects of sustainability are not taken into account at all.
For this reason, among others, this work presents a proposal that solves the mentioned problems.

\section{Approach}\label{approach}
Requirements are a prerequisite to building an application. 
Therefore, we first introduce the goals followed by the application's requirements. 
The actual implementation of the containing spaces is described as a result of the goals and the concrete requirements. 
Thus the application can be evaluated and validated based on the set requirements.  

\subsection{Requirements}\label{requirements}
Among other things, requirements specify and constraints the functionality of an application. 
Therefore requirements are of value for other managers or developers who also want to adapt the concepts of this work to their implementation or existing system. 
Another advantage is that stakeholders can easily verify requirements, and a further evaluation of the system can provide meaningful new insights~\cite{4384163}.

An application usually reflects a vision and therefore has some goals. Five domain experts in the purchasing and business development field have been interviewed to figure out the goals of the DPW. They are all employed at Bosch and responsible for the management and/or development of digital products. 
Through the interpretation of these interviews, the visions of the applications could be built, which significantly reflect the following goals.

\begin{enumerate}
    \item Enabling digitalization and acceleration of digital processes that were previously not digital or only partially digital.
    \item Display aggregated data and allow interaction with them.
    \item Allow parallel work and collaboration between colleagues and also departments.
    \item Create knowledge transfer in terms of transparency between departments and colleagues.
\end{enumerate}

Since those goals are difficult to verify, more concrete requirements are inferred. 
The requirements were specified following the template for requirements by \textit{The SOPHISTs}~\cite{TheSophists2016} and are summarized in the sequel:

\begin{enumerate}
    \item The DPW shall be able to show user-relevant data, information (clustered, aggregated, and summarized news), and tasks.
    \item The DPW shall be able to show aggregated data that are not only maintained by the user's department.
    \item The DPW shall be able to allow collaboration with all colleagues that uses the application.
    \item The DPW shall be able to allow working on processes that were not digital before.
    \item The DPW shall be able to provide analytical evaluations in the form of calculations as well as through visual representations.
    \item The DPW shall be able to provide clustered as well as searchable and filterable data.
    \item The DPW shall be able to be personalized for every user.
    \item The DPW shall be able to break down the supply chain for all relevant products.
    \item The DPW shall be able to offer potential new insights based on analyzed data.
    \item The DPW shall be able to automate low-risk processes.
\end{enumerate}

\subsection{Implementation}\label{implementation}

We implemented the requirements in an application that includes three main concepts: \textit{Procurement Information Space}, \textit{Procurement Analytics and Automation Space}, and \textit{Sustainable Sourcing Space}. 
To enable these three spaces, the application consists of multiple import jobs that fetch data from different data silos. 
Those silos can also be from other departments and are not only maintained by the purchasing department. 
The imported raw data is aggregated, analyzed, and interpreted before being stored in a single database. In this way, we can fetch data faster, and reduce the number of on-the-fly calculations. 

The User Interface (UI) handles user logins via Single-Sign-On and then allows the fetching of user-relevant, supplier-relevant, and/or material group\footnote{A material group, e.g., screws and springs, categorizes materials with the same attributes.}-relevant data.
This way, the separation of concerns is granted.
Other views are, for example, the news feed and the admin dashboard. 

\begin{figure}[htbp]
     \includegraphics[width=0.45\textwidth]{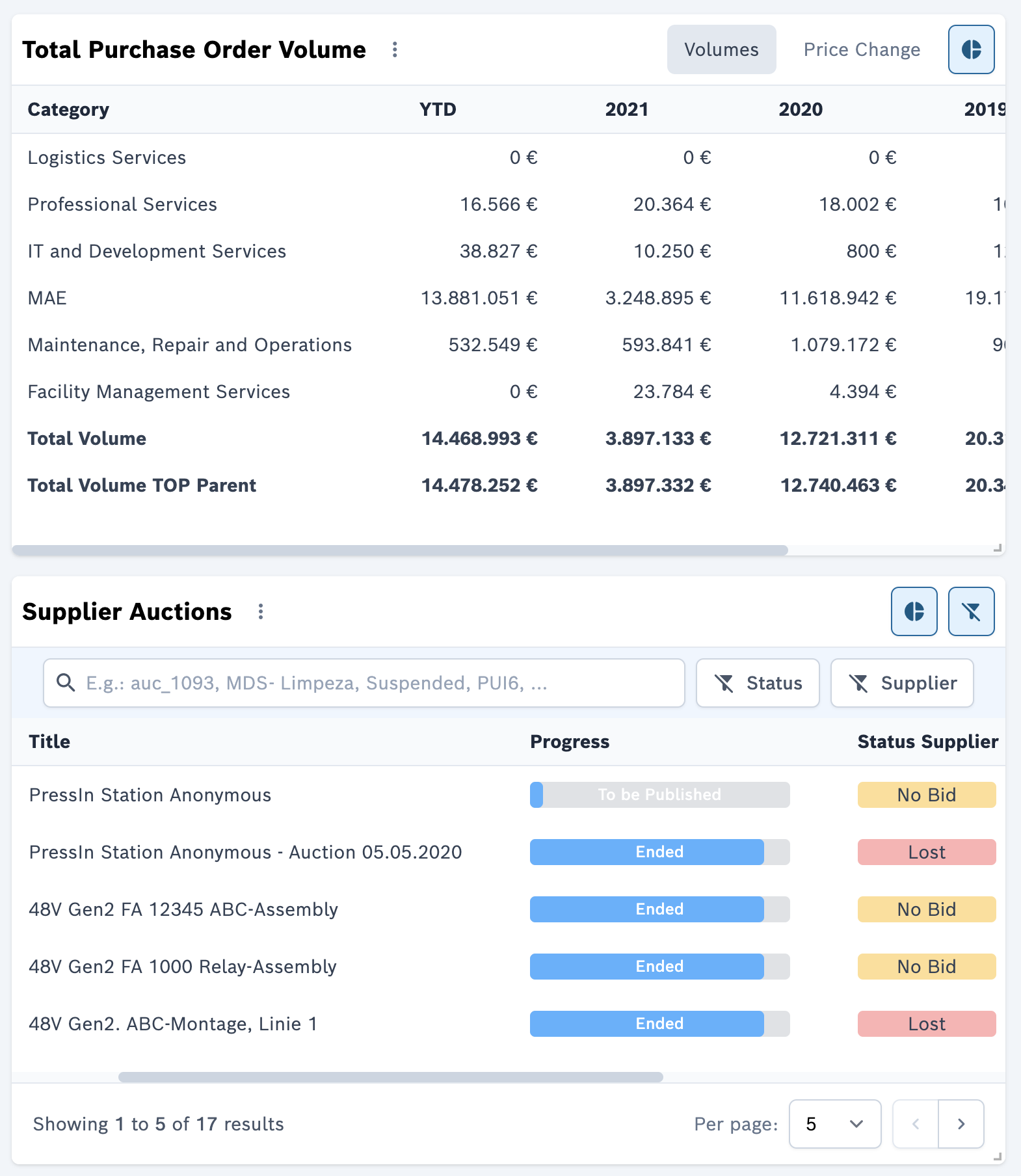}
    \caption{Overview of open auctions of a purchaser and the total purchase order volumes in Euro.}
    \label{fig:dpw_home}
\end{figure}

\begin{figure}
    \includegraphics[width=0.45\textwidth]{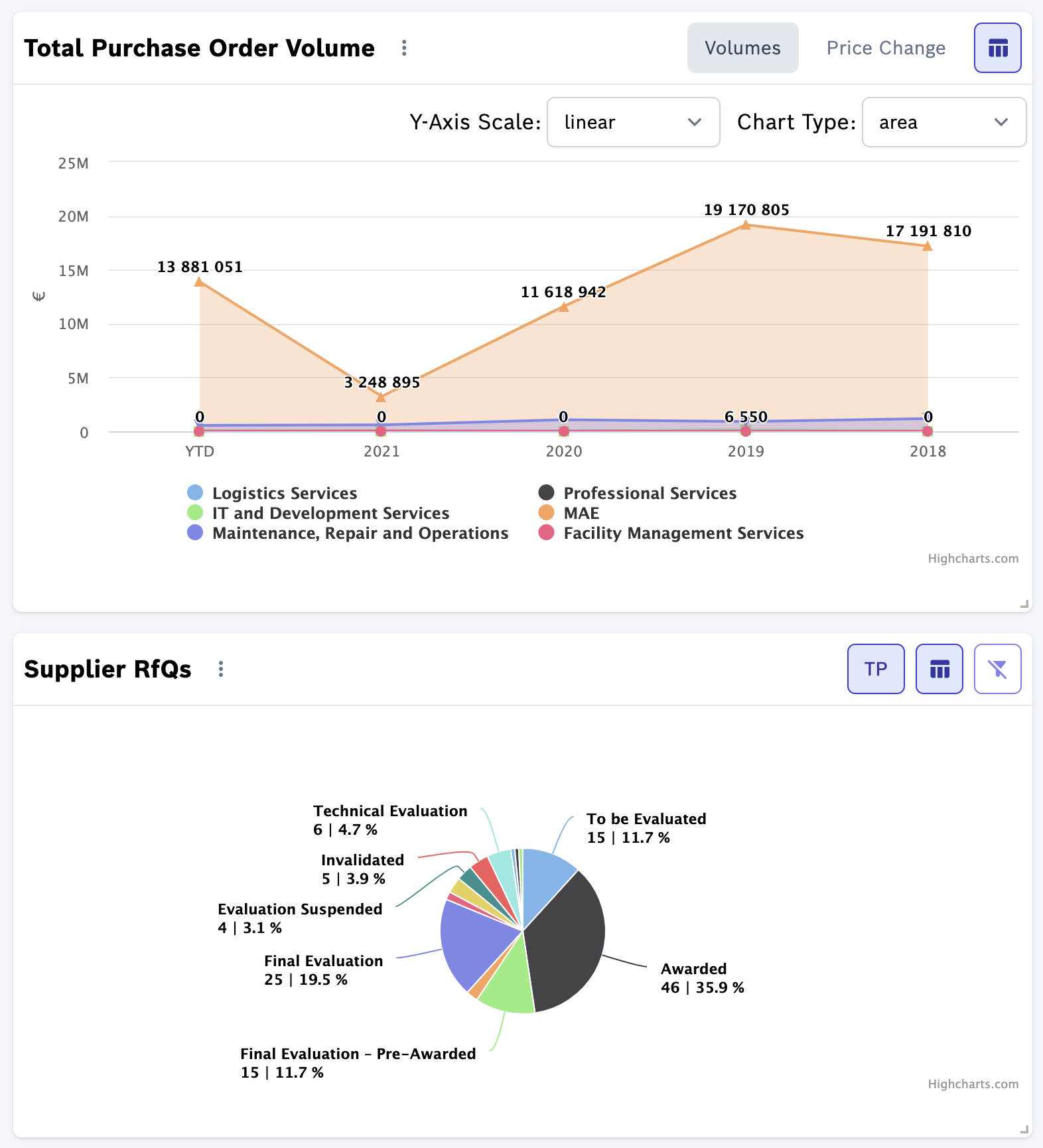}
    \caption{Overview of total purchase order volumes (top), and all RfQs related to the selected supplier (bottom).}    \label{fig:dpw_supplier}
\end{figure}

\begin{figure*}[htbp]
    \centering
    \includegraphics[width=0.85\textwidth]{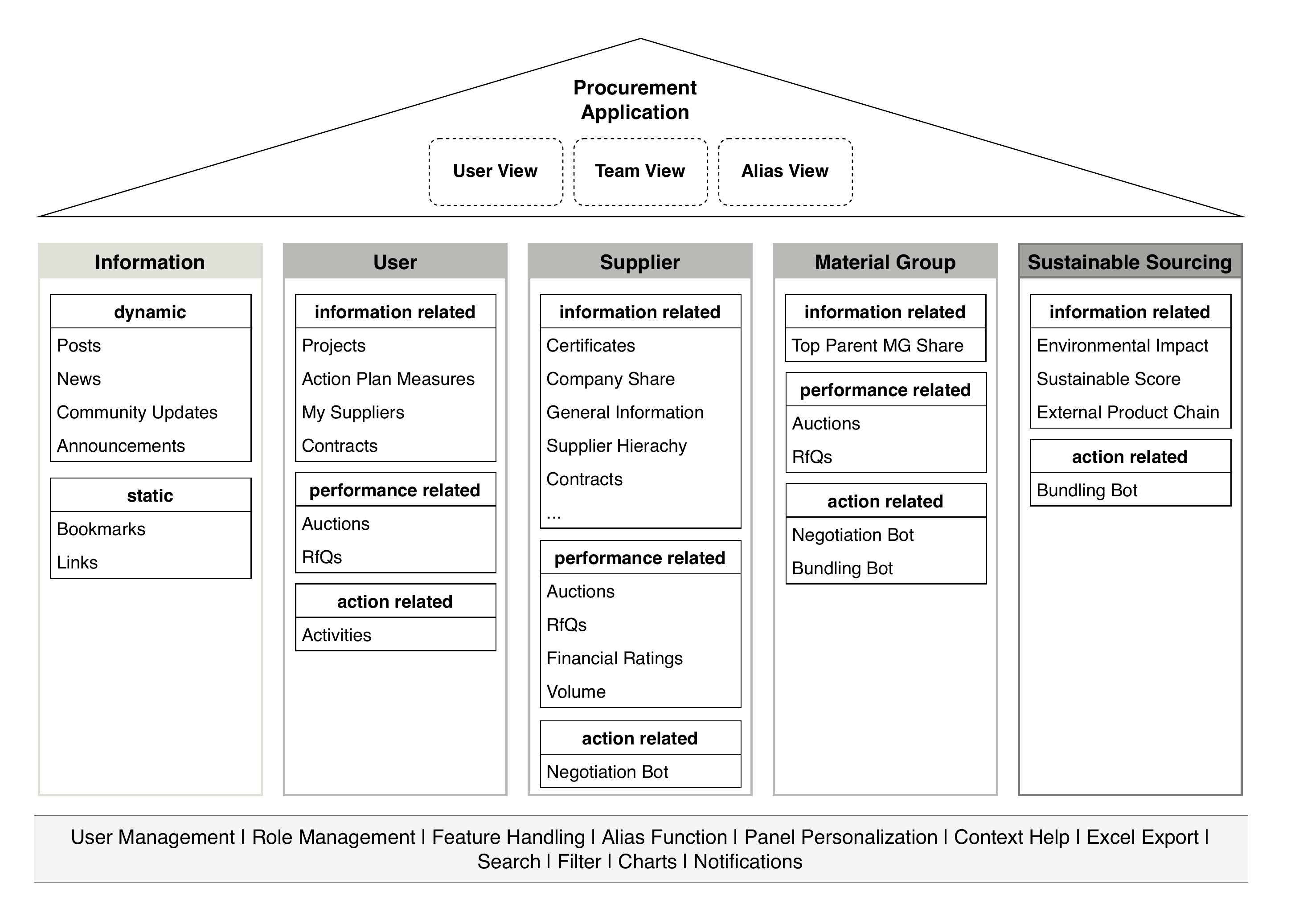}
    \caption{This figure illustrates the Digital Procurement Workspace. The left pillar represents the PIS, the three in the middle the PAAS, and the right one the SSS. The top indicates the different views. The bottom lists all fundamental functionalities.}
    \label{fig:pillar-model}
\end{figure*}

Besides the main features contained and described in the corresponding spaces, more minor features will only be mentioned and not described in detail. 
Those are, for example, filtering and searching of data, suppliers, material groups, and users, favoring news, suppliers, and links, downloading tables, and showing context help for new hires. 
It is possible to adjust the application's layout in widget sizes and positions to allow a personalized view. 
Additionally, users can switch between the user-, team\footnote{The team-view also shows data from team members.}-, and alias-view\footnote{The alias-view allows seeing data from other users.}.
Some widgets can be seen in \autoref{fig:dpw_home} and \autoref{fig:dpw_supplier}. \autoref{fig:pillar-model}, on the other hand, shows the implemented features of each space, the different views, and the minor functionalities in a five-pillar model.

\subsubsection{Procurement Information Space}
The Procurement Information Space provides all needed information regarding news, the latest posts, community updates, and announcements. 
It represents a central point of information gathering and aims to keep users up-to-date and well-educated in their profession. 
Various structured and unstructured data sources are accessed to automatically collect, aggregate, and summarize their content to provide all the user- and purchasing-relevant information.
After an automatic summarization, the information is clustered and provided in natural language to a user. 
Even though there can be different data sources with nearly the same content, those are considered and handled the same way to ensure the validity of the information so the PIS can act like a single point of truth. 
Data sources for news, posts, community updates, announcements, and drop-off points to other tools are configurable and provide even more information. 
While interacting with the features of the PIS (for example, the news feed), the information gets more and more personalized but also considers purchasing-relevant news based on team members' suggestions and reading history. 
The PIS aims to improve decision-making and time-saving as new information could lead to better and faster decisions.

\subsubsection{Procurement Analytics and Automation Space}
The PAAS allows users to access, analyze and interact with procurement-relevant data like open RfQs, auctions, and many more. All that data is divided into three focus groups:

\begin{enumerate}
    \item \textit{User-relevant data} like open RfQs, auctions, and contracts of the logged-in user
    \item \textit{Supplier-relevant data} like RfQs, auctions, and contracts belonging to the selected supplier
    \item \textit{Material group-relevant data} like RfQs and auctions for the selected material groups
\end{enumerate}

The structured data is mainly displayed in a table view, as can be seen in \autoref{fig:dpw_home} (widget names ``Total Purchase Order Volume" and ``Supplier Auctions"), but for a faster overview of some widgets, a chart view is the default view, as can be seen in \autoref{fig:dpw_supplier} (``Total Purchase Order Volume" and ``Supplier RfQs"). 
The chart view naturally cannot provide the same richness of details as the table view, but highlights the most important values, like statuses. 
In contrast, the table view additionally provides further interaction. In this way, procurement-relevant data is updatable from the responsible user.
Besides a graphical summarization of procurement-relevant data, users can also trigger procurement-relevant bots in trivial and/or low-risk tasks and processes. 
Those aim, for example, to reduce human-to-human interaction regarding negotiating low-risk volumes of materials. 
Another use case is the bundling of RfQs across different departments. 
Using these and other bots saves time and money since actual work is automatized and volumes measured in Euro are optimized. 
Other features of the PAAS support the decision-making of purchasers since they provide, for example, a supplier rating based on different well-maintained characteristics, a forecasting of purchasing volumes, and a material group share that indicates the share of each supplier for the selected material groups. 
Improving transparency is achieved by breaking down complex cross-department processes while highlighting the current state of the processes and the actual task logged-in users have to fulfill. 

\subsubsection{Sustainable Sourcing Space}
Transparency regarding environmental and risk impact, as well as fostering a sustainable supply chain, is the main subject of the SSS.
Therefore, this space monitors the environmental and climate impacts of actual materials, products, and suppliers.
In this way, purchasers can directly see the impact of their sourcing decision. 
The SSS automatically gathers information and data on suppliers and their products from internal and external sources to calculate a transparent so-called sustainable score. As a result, we can reduce manual research of purchasers and supply chain experts.
This sustainable score can trigger alerts since a potential risk in the supply is detected or suggests materials or products from different, more sustainable suppliers.
Generally, the sustainable score exists in four stages but concentrates either on Corporate Carbon Footprint (CCF) or Product Carbon Footprint (PCF). 
The first stage is the monetary CCF approach. In this approach, the CO2 emissions are obtained based on the responsible revenue of a supplier. 
For example, if the Anonymous Company is responsible for 10\% of a supplier's revenue, it also obtains 10\% of its CO2 emissions. 
The second CCF approach is based on the determined CO2 emissions for different sectors done by a third party. 
So external databases are accessed to retrieve data about the CO2 emissions if, e.g., steal-related products are purchased from a specific supplier. 
The third approach is also based on third parties and external databases. 
There, the determined emissions for specific products are taken from such databases. 
In this third stage, the PCF is calculated, which also applies to the last stage. 
There, the actual emissions measured by the supplier are communicated and taken for further CO2 emissions reporting. 
In general, there is a focus only on CO2 emissions. 
Other environmental- and climate-relevant values are represented in CO2 equivalents. 
As a result of the sustainable score and the entire SSS, the supply chain should be more resilient, less prone to stoppages, and more environment-friendly.
This way, the goals for the overall purchasing strategy can be achieved more quickly, and the time for finding a fitting supplier is reduced. 
Additionally, the emissions of current products can be reduced, the company's reputation can be increased, and, as Valbuena and Mandojana~\cite{valbuena2022encouraging} describe, effective and sustainable strategic partnerships can be achieved.

\section{Evaluation}\label{evaluation}
We conducted several interviews to evaluate the developed application. 
The interviews aimed to analyze whether the requirements and overall goals were fulfilled, and the expected benefits were obtained. 

\subsection{Study Design}
To verify that the introduced requirements are correctly implemented in the application, qualitative and quantitative analysis in the form of expert interviews has been done. 
The interviews were held after the launch of the first version of the DPW.
The experts have been asked about their experiences a purchaser makes while interacting with the developed application that contains the described concepts.
First, they are asked about the benefits of each concept.
Then the interview changed to a quantitative manner, and the experts could choose between different predefined answers regarding the time savings for each concept.
The same has then been done regarding decision-making support. 
After that, the interviews changed again to a qualitative manner to summarize the main benefits of the application and to give the experts a chance to say anything about the DPW the interviewer had not asked for.

\subsection{Sample}
In total, ten experts were interviewed. 
Two of them are women, and eight are men. 
The experts are all employed at Bosch and have different job roles like product owner, department lead, sustainability expert, junior and senior purchaser, and innovation manager. 
The interview showed that most experts could answer mostly without any counter questions.
Nine of the ten interviews were valid, and all interviewees had the same interviewer and the same interview style, and all interviews were conducted online, while nine were held in German and one in English.
One interview could not be considered since the interviewee struggled with answering the questions and wanted to avoid the interview in general.

\subsection{Data Analyses}
The interview transcripts were, according to Mayring \cite{Mayring_2000}, answer-by-answer summarized, analyzed, and further categorized. 
This way, various categories have been found. 
Using Mayrings methods, the summarization continues until statements can be generalized and further clustered so that categories can be combined~\cite{Mayring_2000}. 
As a result, each category is defined by a definition statement. 
To compare the categories in their importance, they have been counted. 
The description of the complete coding scheme is beyond the scope of this paper, but the emerging categories are presented in the results section.
The quantitative part of the interview is handled differently. 
The answers to each question regarding time-saving and support in decision-making are counted and presented in the next subsection. 

\subsection{Results}
In total, eleven categories describing a DPW's benefits are found. They are listed and further described in \autoref{table:benefit_frequency} and published in their raw data on Zenodo~\cite{jan_david_stutz_2022_7323210}.
During the interviews, some experts mentioned several shortcomings and risks. Since we have not explicitly asked for shortcomings or disadvantages, those are further discussed and interpreted in \autoref{discussion}.

\begin{table*}[!t]
\small
\caption{Overview of categories mentioned across all three spaces, with definitions and examples.}
        \begin{tabular}{ |p{1.7cm}|p{1cm}|p{3.5cm}|p{6.5cm}|p{1.3cm} | } 
        \hline
        \textbf{Category} & \textbf{Frequ-ency} & \textbf{Definition} & \textbf{Example} & \textbf{Require-ment}\\ 
        \hline
        Transparency & 14 & Statement that indicates an increase in transparency & ``If we want to achieve a certain target, in terms of CO2 footprint, then of course that relates to the entire supply chain." & 2, 8, 9  \\
        \hline
        Efficiency & 11 & Statement that efficiency in time and the financial outcome is increased. & ``You can organize activities in a more targeted manner and thus work more efficiently, of course, and you can also improve processes and then lower prices." & 4  \\
        \hline
        Decision-Making & 9 & Statement that the process of decision-making is improved and better decisions are made. & ``It brings together decision-relevant information quickly and compactly so that the buyer makes the right decisions or is supported in the decision-making process." & 9 \\
        \hline
        Centralization & 8 & Statement that information and tasks can be worked in one central application. & ``You have personalized information in one place without system and media discontinuities." &  1, 2, 3   \\
        \hline
        Optimization & 8 & Statement that the application optimizes the workflow of purchasers. & ``Now you see: where I haven't had a price change for a long time, how have my raw material prices developed, is this a good time to negotiate, or a bad time to negotiate." &  1, 2, 5, 6, 9, 10 \\
        \hline
        Strategy & 7 & Statement that the company or department-wide strategy is adapted or adhered to. & ``If I also have a real conflict of goals as a result and I notice the effects on my subgoals more strongly, then I probably decide differently." & 3, 9  \\
        \hline
        Time-saving & 7 & Statement that actual time is saved. & ``If the buyer today somehow has to do a lot of recurring, i.e. relatively monotonous, tasks that add zero value, this naturally ties up capacity that is now freed up." & 4, 10 \\
        \hline
        Personal-ization & 5 & Statement that the application provides personalized information and is still further adjustable. & ``You have the information straight to you, they are Tailor-Made for you." & 1, 7 \\
        \hline
        Insights & 4 & Statement that new insights are gained. & ``Purchasers don't really know 80 to 90 percent of their suppliers. This means that if there is potential there, you have to actively point it out to them." & 9 \\
        \hline
        Collaboration & 3 & Statement that the collaboration and communication of employees are increased. & ``Today, this is used not only for purchasing, but also for other areas such as logistics, sales, and production, so that everyone can talk directly to each other." & 3 \\
        \hline
        Informed & 3 & Statement that employees are in general better informed. & ``This also ensures that everybody is at the same level of information."       & 1  \\            
        \hline
        \end{tabular}
        \label{table:benefit_frequency}
    \end{table*}

\begin{figure}[htbp]
    \includegraphics[width=0.475\textwidth]{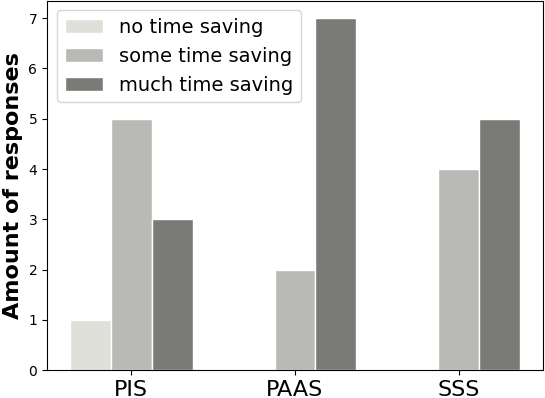}
    \caption{Number of answers regarding time saving.}
    \label{fig:time-evaluation}
\end{figure}

\begin{figure}[htbp]
    \includegraphics[width=0.475\textwidth]{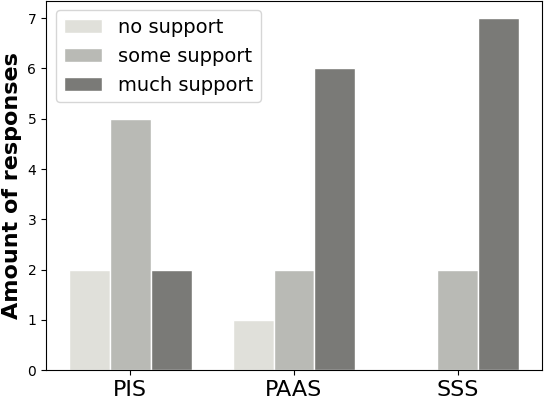}
    \caption{Number of answers regarding support.}
    \label{fig:decision-evaluation}
\end{figure}

For the three concepts, the weighting and occasionally the benefits as a whole differ. 
The PIS mainly increases the purchasing experience regarding scouting new suppliers and day-to-day workflows, resulting in increased efficiency, which five of the interviewed experts noted. 
Increased efficiency for a purchaser means more suppliers can be contracted manually and with better preparation which, based on the experts, leads to better results in Euro.
 
On the other hand, seven of the experts noted that PAAS mainly leads to optimization, while six experts see increased efficiency in, for example, sourcing, negotiation, and contracting. 
However, it also provides increased transparency and leads to time savings for purchasers. 
Transparency is seen as the SSS's main benefit, which eight of the experts noted. 
However, the space also increases decision-making and adherence and further development to the central purchasing strategy. 
Some benefits of the single spaces had only a few mentions. 
Nevertheless, some of them are highlighted as the main benefit of an application that contains such concepts. This fact applies especially to the collaboration aspect. 
While the respective spaces miss basic features regarding collaboration, experts mentioned collaboration as one of the main benefits of the application. 
In general, the main benefits are increased transparency and time saving, better decision-making, an increased collaboration of purchasers and departments, and a higher purchasing experience in day-to-day work through personalizing the entire application containing the three spaces.

The quantitative part of the interview focused on decision-making and time-saving since those two can be seen as the main advantages of business information systems and are also mentioned as the main goals of the domain experts interviewed while setting up the application requirements~\cite{mesarovs2021impact}. 
During the evaluating expert interviews, the participants were asked if the respective space (PIS, PAAS, SSS) does not save time, saves some time, or saves much time. 
The same was done regarding decision-making, while the predefined answers were no support in decision-making, some support in decision-making, and much support in decision-making. 
As the evaluation in \autoref{fig:time-evaluation} and \autoref{fig:decision-evaluation} shows, most time-saving can be achieved by implementing a feature-rich PAAS. 
Based on the experts, decision-making is supported almost equally while implementing a PAAS and SSS. 
Nevertheless, PIS is considered in both categories as some support.

\section{Discussion}\label{discussion}
One of the goals of this work is to evaluate the application against the set requirements. 
The expert interviews were designed to verify the requirements. 

\subsection{Verification of the Set Requirements}
All requirements introduced in \autoref{requirements} are fulfilled due to the implementing of the concepts.

User-relevant information is displayed via the PIS, and tasks can be assigned inside the PAAS. The experts also mentioned as a benefit that due to these spaces, users are more informed, and the workflow is considered more optimized. 
Thus, the first requirement is fulfilled.
Assigning tasks to others and commenting on processes or RfQs leads to increased and optimized collaboration. This was indicated by the experts directly and is seen as an overall benefit of the developed application while fulfilling requirement three. 
On the other hand, requirements one and seven are fulfilled by favoring suppliers, selecting the news, and adjusting the application's layout. That leads to time savings and a personalizable application. 
Furthermore, the implementation of the PAAS provides the interaction (filtering, searching, adjusting) of aggregated data from different data sources. 
Those data sources are partially maintained by different departments. 
The aggregated and analyzed data allows further evaluations and visual representations. 
According to the experts, this results in an optimized application that improves decision-making, increases efficiency, and saves time. 
Thus requirements one, two, five, six, nine, and ten are fulfilled. 
In some cases, processes that were not digital before are automatable using bots. 
Besides, some triggerable bots can also automate low-risk processes and thus reduce human-to-human interaction. 
Those bots save time and are responsible for fulfilling requirements four and ten.  
The SSS offers to break down the supply chain and provides further insights regarding the sustainability, risk, and environmental impact of single materials and products. 
According to the experts, that leads to increased transparency and new insights, fulfilling requirements two, eight, and nine. 
Furthermore, the increased transparency and the new insights can also lead to strategy adjustments, which can directly reflect carbon emissions. 

Among others, most business information systems have the goal of reducing costs and time~\cite{mesarovs2021impact}. 
Either while saving time for employees or improving decision-making so the company can reach a better result in Euros. 
By implementing the introduced concepts, the DPW saves time and improves decision-making. 
Besides a company's performance, climate change and the goal of reaching the set goals of the Paris Agreement for 2050 are additionally relevant. 
Since the experts mentioned that the enrollment of the DPW improves the company's performance but also fosters procurement regarding sustainability and resilience, the overall goals can be seen as achieved, even though not all relations of the found benefits and the created requirements are described here in detail.

\subsection{General Findings}
Besides verifying the set requirements, other meaningful new insights are found, like dependencies of benefits, some risks of the application, and suggestions for potential enablers of such a DPW.

In general, it is recommended to focus more on the PAAS and SSS.
\autoref{fig:time-evaluation} shows that according to the interviewed experts, the most time can be saved by implementing the PAAS concept, while SSS could become more and more relevant in the future.
On the other hand, \autoref{fig:decision-evaluation} shows that SSS leads to better decision-making. 
Due to the increased impact of purchasers' daily business regarding sustainability and the environment and studies that have already shown that green digital procurement applications foster Big Data analysis competencies, it is recommended to set fundamentals for SSS early in development~\cite{AlNuaimi2021}. 
Concretely it is recommended to start with centralizing information and allow as well as support personalization features from the beginning of development since they influence time-saving and decision-making and are independent of other categories. 
After centralizing and during the development of new features, keep in mind that the new features should either improve or support transparency (e.g., showing the supplier's responsible), collaboration (allowing assigning tasks), or optimization (aggregate data).
As a result, new insights should occur, and the possibility to adapt the purchasing strategy based on the new insights is given.
To sum it up, if done right, a centralized and personalizable application leads to the mentioned time-saving, higher efficiency, and improved decision-making but enables, in general, all other benefits presented in \autoref{table:benefit_frequency}. 

The DPW generally provides more features than the presented competitors from \autoref{procurement_apps}. 
Especially SSS- and PIS-related features are entirely missing in those applications.
While introduced functionality of PAAS also lacks in those applications. 
The found benefits of the DPW are also not transmittable to the earlier described applications. 
Moreover, it is doubtful that the found and introduced benefits can also be achieved in those applications. 
Therefore, implementing a customized DPW is a good solution.

\subsection{Dependencies and Risks}
\begin{figure}
    \includegraphics[width=0.45\textwidth]{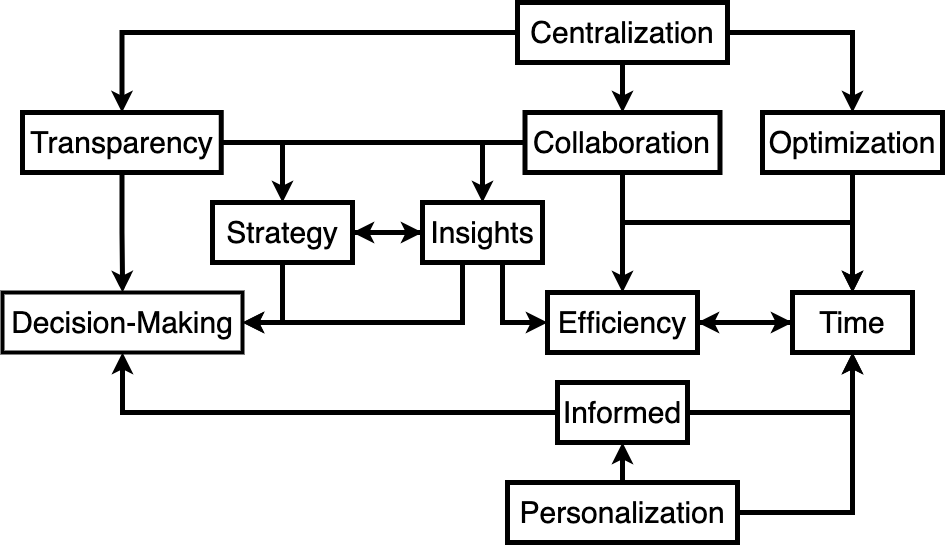}
    \caption{Influences of the different categories. The arrow direction indicates the influence direction.}
    \label{fig:dependencies}
\end{figure}

While evaluating the results, we discovered some dependencies of the found benefits presented in \autoref{table:benefit_frequency}. 
In fact, some benefits act as a prerequisite for others. 
For example, \textit{Centralization} is considered and also actually mentioned as the prerequisite for enabling \textit{Transparency} and \textit{Collaboration}. 
That, on the other hand, leads to new insights, which further leads then to improved decision-making and time-saving. 
Another example, according to the experts, is that personalization regarding news, blog posts, and announcements lead to more up-to-date users, resulting in better decisions.
All of those dependencies are outlined and contextualized in \autoref{fig:dependencies}.

While centralization is considered a prerequisite for most other benefits, as seen in \autoref{fig:dependencies}, it is also mentioned as a risk for the company since a central application could lead to a single point of failure. 
Besides that, increased feature richness could lead to complex maintenance, which results in higher costs that purchasers must balance with a more efficient procurement. 
Nevertheless, initial high costs and the maintenance for developing applications and new features inside the application are generally known issues in service-oriented software applications.

Due to too many powerless features with a poor User Experience (UX), users could start to work less efficiently or, worst case, avoid the application. 
To prevent unsatisfied users, it is helpful to educate them, either with software or through personal training. 
Another suggestion is to allow user inputs as early as possible because if that is not the case, users must use different additional tools for updating data, which leads to decreased user acceptance and satisfaction.

\subsection{Future Work}
The evaluation of the developed application focuses mainly on the benefits for the company itself and its positive impact on the users. 
Future work should also consider the disadvantages of such an application. 
Even though centralization and the lack of proper UX are mentioned, experts can deliver even more possible disadvantages if they are specifically asked for them.  

Today the entire application does not use a knowledge graph architecture. 
Currently, all the data is imported from different sources and stored in a central database. 
Ma and Moln{\'a}r~\cite{Ma2019} suggest using ontologies and a knowledge graph because that is considered an effective technology to integrate data from multiple heterogeneous sources. 
Therefore, future work should use a knowledge graph as the underlying technology while implementing the presented concepts and features. 
Even if an application implements the presented features based on knowledge graphs (or any other underlying technology), there is still a risk left. 
Poor UX could lead to unsatisfied users, and as Magnus and Rudra~\cite{Magnus2019} claim, dashboards built based on principles of cognition enhance decision-making in a supply chain. 
Therefore, future work should focus more on UX-related topics. \\

\section{Conclusion}\label{conclusion}
Due to the urge to accomplish carbon neutrality, wars, sanctions, the pandemic, and catastrophes, sustainable SCM has become even more important for companies. 
As a global Fortune 500 company, we have developed an application that encompasses a novel way of information integration, automation tools as well as analytical techniques. 
All developed features can be categorized into one of the three introduced concepts - the Procurement Information Space, the Procurement Analytics and Automation Space, and the Sustainable Sourcing Space. 
In-depth expert interviews were conducted to verify that the implemented spaces perform as expected. 
The results of the interviews show that the set requirements are met. Besides that, the interviews revealed increased time-saving and support in decision-making, especially due to the Procurement Analytics and Automation Space and the Sustainable Sourcing Space. 
Increased transparency, efficiency, and decision-making support are mentioned most as a benefit, but they may not be the most important ones. 
This work encourages other purchasing departments to build their own DPW and provides valuable suggestions on what concepts and features a DPW should implement. 
Existing applications can also profit from this work since new concepts are presented, and the evaluation of the benefits could lead to strategy adjustments. 
Such strategy adjustments could then influence procurement regarding a more sustainable, resilient, and more carbon-reduced supply chain so that a Digital Procurement Workspace can help reach the goals of the Paris Agreement in the next twenty-seven years. 

\section*{\uppercase{Acknowledgements}}
We thank all participants from the Robert Bosch GmbH who were, in whatever case, involved with the Digital Procurement Workspace and this work.
\vspace{-0.5cm}

\bibliographystyle{apalike}{
    \small
    \bibliography{example}
}

%\section*{\uppercase{Appendix}}

% \textbf{Expert Interview Questions}

\end{document}